\documentclass[11pt,twoside]{article}
\usepackage{asp2014}

\aspSuppressVolSlug
\resetcounters

\begin{document}

\title{Difference frequencies of nonradial pulsation modes 
\\
and repetitive outbursts in classical Be stars\footnotemark}
\author{D.\  Baade$^1$, Th.\ Rivinius$^2$, A.\ Pigulski$^3$, 
A.\ Carciofi$^4$, and BEST$^5$}
\affil{$^1$European Organisation for Astronomical Research in the Southern Hemisphere, 
Garching b.\ M{\" u}nchen, Germany; \email{dbaade@eso.org}}
\affil{$^2$European Organisation for Astronomical Research in the Southern Hemisphere, Santiago, Chile; \email{triviniu@eso.org}}
\affil{$^3$Instytut Astronomiczny, Uniwersytet Wroc\l{}awski, Wroc\l{}aw, 
Poland; \email{pigulski@astro.uni.wroc.pl}}
\affil{$^4$Instituto de Astronomia, Geof{\' i}sica e Ci\`encias Atmosf\'ericas, Universidade de S{\~ a}o Paulo, S{\~ a}o Paulo, Brazil; \email{carciofi@usp.br}
\affil{$^5$ Bright Target Explorer (BRITE) Executive Science Team (BEST)}
}

\paperauthor{Dietrich Baade}{dbaade@eso.org}{}{European Organisation for Astronomical Research in the Southern Hemisphere}{}{Garching b. M{\" u}nchen}{}{85521}{Germany}
\paperauthor{Thomas Rivinius}{trviviniu@eso.org}{}{European Organisation for Astronomical Research in the Southern Hemisphere}{}{Santiago}{}{}{Chile}
\paperauthor{Andrzej Pigulski}{pigulski@astro.uni.wroc.pl}{}{Instytut Astronomiczny, Uniwersytet Wroc\l{}awski}{}{Wroc\l{}aw}{}{Postal Code}{Poland}
\paperauthor{Alex Carciofi}{carciofi@usp.br}{}{Instituto de Astronomia, Geof{\' i}sica e Ci\`encias Atmosf\'ericas, Universidade de S{\~ a}o Paulo}{}{S{\~ a}o Paulo}{S{\~ a}o Paulo}{SP-05508-900}{Brazil}
\paperauthor{BRITE Executive Science Team (BEST)}{}{}{}{}{}{}{}{}

\titlefootnote{Based on data collected by the BRITE Constellation
satellite mission, designed, built, launched, operated and supported by
the Austrian Research Promotion Agency (FFG), the University of Vienna,
the Technical University of Graz, the Canadian Space Agency (CSA), the
University of Toronto Institute for Aerospace Studies (UTIAS), the
Foundation for Polish Science \& Technology (FNiTP MNiSW), and National
Science Centre (NCN).}
\begin{abstract}
Space photometry has revealed rich pulsation spectra in classical Be
stars.  Often frequency pairs can be found that differ by nearly the 
frequency of a third genuine stellar variability.  
The lowest currently known of these so-called difference (or $\Delta$) 
frequencies reach 0.03 c/d, which is the upper limit of outburst frequencies.  
The potential of
such slow variations for pulsation-assisted mass loss is discussed.
\end{abstract}

\section{Introduction} Classical Be stars pose two core problems: (i)
How do they eject matter?  (ii) How can ejecta form a Keplerian disk?
As Ghoreyshi et al., Klement et al., and Vieira et al.\ illustrate in
these proceedings, viscosity is the key process that drives the
exchange of angular momentum between gas parcels so that $\sim$1\% of
the matter reaches an orbit.  By contrast, the first Be-star challenge
has not yet found a well-documented general solution.  The only case
of clear causality was presented by \citet{1998cvsw.conf..207R}, who
observed that outbursts of $\mu$ Cen occur when two of six
spectroscopic nonradial pulsation (NRP) modes are in phase at the
stellar surface.  When ranked by velocity amplitude, both the first
and the second as well as the first and the third mode combine in this
fashion.  Very plausibly, it was interpreted as a beat phenomenon
although direct evidence of a typical beat pattern was not produced.
\citet{2001A&A...369.1058R} identified all three as $\ell = m = 2$
$g$-modes which prevail in most Be stars \citep{2003A&A...411..229R}.

This relatively thin factbase was widely adopted and modified to the
more general notion that during outbursts of Be stars an unspecified
number of NRP modes somehow combine to liberate the energy needed to
eject significant amounts of matter
\citep[e.g.,][]{2009A&A...506...95H, 2015MNRAS.450.3015K}.  This
conclusion supposedly received support from photometric
observations of Be stars in outburst when many spikes in the power
spectra change their strength, with a large fraction of them becoming
detectable only around the outbursts.  To this end, the spikes were
interpreted as NRP modes.  However, \citet{2016A&A...588A..56B} and
\citet{2016arXiv160802872R} questioned this identification
and showed that much of the variability power involved arises from
circumstellar processes.

\section{Space photometry}
Spectroscopic monitoring of broad-lined stars pulsating in multiple
modes and undergoing additional outbursts is a very resource-demanding
undertaking.  However, not only is the technique expensive, but in about
20 years it has found only that one case of $\mu$ Cen.  28 Cyg and $\eta$ Cen
were eventually confirmed by photometry (cf.\ Sect.\ 3).

The advent of photometry with tiny to medium-sized
satellites and corresponding costs has opened up new opportunities
with the amenities of space observatories: 
SMEI \citep{2013SSRv..180....1H}, 
MOST \citep{2003PASP..115.1023W}, 
CoRoT \citep{2006ESASP1306...33B},
Kepler \citep{2010PASP..122..131G}, 
and 
BRITE \citep{2016arXiv160800282P}.  
All five missions
have observed Be stars.  Examples (except for MOST) will be discussed
in the next section.  

Even precision photometry is not perfect: In rapid rotators such as Be
stars, frequencies are subject to large rotational modification and
can be treacherous indicators of mode types.  Moreover, the
latitudinal component of low-order $g$-mode velocity fields is a
strong identifying symptom in spectra of pole-on stars whereas the
photometric signal is largest in the equatorial plane but not visible
in true pole-on stars. In Be stars, there is an additional photometric
pitfall: Light from the central star is reprocessed in the disk.
Therefore, variable amounts of circumstellar matter contribute an
additional variability by this kind of light echo, which is weakest at
inclination angles around 70 degrees \citep{2012ApJ...756..156H}.  A
strong example is $\mu$ Cen, where this effect precluded the detection
by BRITE of the spectroscopic modes \citep{2016A&A...588A..56B}, the
photometric signature of which is low at the given low inclination.

\section{Difference (or $\Delta$) frequencies}
In the following, frequency differences and difference (= $\Delta$) frequencies
need to be carefully distinguished: The former can be calculated
between any two frequencies.  A $\Delta$ frequency is something much
more specific because, in addition to being a simple frequency
difference, it plays its own role as a third variability.  This is not
the case for arbitrary frequency differencies, which do not normally
have any physical meaning; in power spectra of synthetic data with 
two sinusoidal variations, difference (and sum) frequencies do not occur.  
$\Delta$ frequencies were found in the 
following Be stars:

\noindent
{\bf $\eta$ Centauri} (B2):
Two retrograde $\ell=2$ $g$-modes with 1.732 and
1.764 c/d are known from spectra \citep{2003A&A...411..229R}.  In
agreement with the nearly equator-on orientation and the low H$\alpha$ 
line-emission activity, BRITE detected both with relatively large 
amplitudes slightly above
2 mmag \citep{2016A&A...588A..56B}.  A frequency of 
0.034 c/d roughly equal to the difference between the two NRP
frequencies was also discovered.  It has two unexpected properties: At
15 mmag its amplitude is 2.5 times as large as the sum of the two NRP
modes, perhaps indicative of nonlinear mode coupling.  And it is
not a beat frequency but an approximately sinusoidal variability in
its own right.  Increased scatter around the extrema of the associated
light curve and modifications of the amplitude and phase of the
circumstellar {\v S}tefl frequency \citep{2016A&A...588A..56B} with the
stellar $\Delta$ frequency imply that the latter
modulates and perhaps even drives mass loss.

\noindent
{\bf 28 Cygni} (B2.5):
Two (numerically imperfect) $\Delta$ 
frequencies occur in BRITE data (Baade et al., in prep.):
0.051 c/d $\approx$ 1.598c/d - 1.544 c/d and 0.217 c/d $\approx$
1.598 c/d - 1.380 c/d. The amplitude (8.5 mmag) of the slower variability 
exceeds those ($\leq$5.5 mmag) of the two
retrograde quadrupole $g$-modes \citep{2000ASPC..214..232T} while in
the second tuple the amplitudes are less different.  Indications of a
link to mass loss exist as in $\eta$ Cen but are much weaker.  Like in
$\mu$ Cen (provided this star does possess genuine $\Delta$ 
frequencies), the two $\Delta$ frequencies have one parent NRP
frequency in common so that only three NRP modes seem involved
in their formation.

\noindent
{\bf HD 50209} (B9):
From CoRoT observations,  \citet{2009A&A...506..125D} suggest
the presence of rotational frequency splitting although the implied
frequency of the zonal mode ($m$=0) would be unusually low (0.108
c/d).  Indications of rotational splitting are not commonly found in
Be stars.  If HD 50209 is such a case, it illustrates that a frequency
difference is not a difference ($\Delta$) frequency: The frequency 
difference is a quantity with high diagnostic potential but does not 
correspond to a separate stand-alone variability.

\noindent
{\bf HD 49330} (B1):
In the last third of the 137-d monitoring period with CoRoT,
\citet{2009A&A...506...95H} detected a 0.03-mag outburst, during which
the two base frequencies, 11.86 c/d and 16.89 c/d, lost part of their
power.  The difference between them appears as a genuine
$\Delta$ frequency at 5.03 c/d (confirmed by Rivinius et al., in
prep.).  In contemporaneous echelle
spectra, \citet{2009A&A...506..103F} determined an azimuthal mode
order $\ell \approx 4$ for the 11.86-c/d variability and $\ell \approx
6$ for the higher-frequency mode.

\noindent
{\bf HD 186567}:
The Kepler data were analyzed by
\citet{2015MNRAS.450.3015K} and \citet{2016arXiv160802872R}.  The
results based on conventional Lomb-Scargle power spectra agree fairly
perfectly.  In particular, the two investigations found a $\Delta$ 
frequency at 0.276 c/d corresponding to 4.010 c/d - 3.734 c/d.
Both of the latter frequencies are the strongest peaks in a group.
Rivinius et al.\ also conducted a wavelet transform.  It revealed the
two amplitudes as time dependent on a scale of $\sim$ 20 days with
variations being antiphased.  In stark contrast, the variability with
the $\Delta$ frequency only showed a slow and perfectly smooth trend 
over the four years of observations; howvwer, this is also favoured 
by the wavelet method. 

\noindent
{\bf ALS 10705}:
\citet{2016arXiv160802872R} analyzed Kepler data 
and found that the frequency differencies 4.77 c/d -- 4.41
c/d and 5.13 c/d -- 4.77 c/d are about equal, namely 0.36 c/d.  There
is a genuine frequency at 0.35 c/d so that the two pairs may not only be a
case of frequency splitting but give rise to genuine $\Delta$ 
frequencies.  In addition, the difference between 1.58 c/d and 0.59
c/d (0.99 c/d) is nearly coincident with the largest-amplitude
variability at 0.98 c/d.  This could constitute a third $\Delta$ 
frequency.

\section{Be stars without (firmly) detected $\Delta$ frequencies}
BRITE observations of several other Be stars were searched for $\Delta$ 
frequencies.  For most of them some
excuses can be construed as to why none was found:
\begin{list}{$\bullet$}{\topsep=0mm\parsep=0mm\itemsep=0mm}
\item
Late-type Be Stars: In agreement with most other cool Be stars, 
no periodic short-term variability 
with semi-amplitude above 1 mmag was seen in $\kappa^1$ Lup (B9.5) 
and $\mu$ Lup (B8).  The 5.9-mmag amplitude (0.6586 c/d) 
in $\nu$ Pup (B8) is, therefore, remarkable but the only variability above 
the detection threshold.  
\item 
Intermediate-type Be stars: Short-periodic variability was also not detected 
in 17 Tau (B6), 23 Tau (B6), and $\eta$ Tau (B7).  At blue wavelengths, 
$\psi$ Per (B5) exhibits 
a large-amplitude (10.3 mmag) variability with 0.0355 c/d, which numerically 
matches the difference between 0.7182 c/d (1.8 mmag) and 0.6829 c/d (1.9 mmag).
Blue and red data are of different quality and quantity 
and do not give the same results.  In fairly limited datasets, $\phi$ And 
(B5; only blue data available) and HR 1113 (B7; only red data available) 
appear single-periodic with 
0.6500 c/d or 1.3085 c/d (1.1 mmag) and 1.488 c/d (2.7 mmag), respectively.  
\item
In limited data, $\omega$ Ori (B3) does not present a significant 
isolated frequency but a frequency 
group \citep{2016A&A...588A..56B} around 1.05 c/d, which apparently 
changed from 2014 to 2015
and perhaps explains the difficulty of establishing a stable frequency 
\citep{2002A&A...388..899N}.  In even sparser data of 
25 Cyg (B3), only one possible frequency was found (1.86 c/d, 1.5 mmag).  
Well-observed other earlier-type Be stars have much richer power spectra as
illustrated in the following.   
\item
48 Per (B3):  Three large-amplitude variations with widely spaced frequencies
were detected in this earlier-type 
shell star (viewed equator-on): 0.0311 c/d (5.4 mmag), 1.074 c/d 
(4.3 mmag), and 2.2594 c/d (4.2 mmag).  The slow variability is not in any 
obvious way related to more rapid variations.  
\item
60 Cyg (B1):  Among the many significant variabilities, one  stands 
out because both its frequency (3.336 c/d) and amplitude (6.4 mmag) are 
large.  
\item
25 Ori (B1): There are two candidate $\Delta$ frequencies with large 
amplitudes: 0.0268 c/d (11.3 mmag) $\approx$ 
2.9530 c/d (2.0 mmag) - 2.9889 c/d (4.8 mmag) and 0.1897 c/d (15.9 mmag) 
$\approx$ 1.6782 c/d (6.9 mmag) - 1.4888 c/d (9.3 mmag).  Separated 
by three months, phases of slow, roughly sinusoidal variations occurred with 
full amplitudes between 200 and 300 mmag.  
\item
$\gamma$ Cas (B0):  In the combined blue and red data (to extend the 
time span) several frequencies were found:  
0.2734 c/d (1.1 mmag), 0.9728 c/d (2.2 mmag), 2.1838 or 2.1950 c/d (0.9 mmag),
and 2.4797 c/d (2.7 mmag).  The latter is $\sim$3 times the frequency 
of 0.8225 c/d derived by \citet{2012ApJ...760...10H}
from single-site ground-based photometry, which BRITE does not detect 
at the 0.6 mmag level.  If the true frequency is the higher one, its 
interpretation as rotational modulation would put other parameters under 
considerable stress.  However, the only frequency seen with 
SMEI is 0.821895 c/d. 
For a Be star, 0.27 c/d is an unusual frequency.  But it is very close 
to the differenc between 2.4797 c/d and 2.1950 c/d. A possible very low 
frequency at 0.0204 c/d (2.9 mmag) requires confirmation.  
\item 
$\delta$ Sco (B0): In red-light observations, this single-lined 
binary showed several slow variations: 0.0937 c/d 
(0.8 mmag), 0.0305 c/d (1.9 mmag), and 0.01402 c/d (1.5 mmag).
The latter one could be the difference between 1.0316 c/d (0.8 mmag) and 
1.0176 c/d (0.9 mmag).  Unfortunately, the blue data are of lower quality 
and not sufficient to support or question the red data.  
\item 
The BRITE
observations of 28 CMa (B2) became available only after this
workshop.  The SMEI data are noisy, and the star is seen at low
inclination.
\item 
27 CMa (B3) is a strong shell star, i.e.\ viewed
through the disk.  At V=4.7 mag it is very faint for SMEI; the delivery
of BRITE data took place after this workshop.  
\end{list}
In summary, this list is inconclusive.  (It is also very inhomogeneous 
because the number of data, the photon noise, and systematic errors 
vary widely.) Very slow variations (0.02 to 0.05 c/d) without 
obvious instrumental or other explanations as artifacts, are 
wide-spread but cannot always be related to more rapid variations.  Some 
time scales reach 30-40\% of the time span of the data and need confirmation.

\section{Discussion} 
\subsection{Commonalities of variabilities with $\Delta$ frequencies} 
With such a small number of examples, which
may not all belong to the group to be defined, and incomplete
information about most of them, the following compilation is not
likely to fully survive critical scrutiny for long but may serve as a
first guidance:
\begin{list}{$\bullet$}{\topsep=0mm\parsep=0mm\itemsep=0mm} 
\item 
In most stars with spectroscopic mode identifications ($\eta$ Cen, $\mu$
Cen, 28 Cyg), the parent variations are retrograde quadrupole
$g$-modes.  The only counter example is HD 49330 with $\ell \approx 4$
and $\ell \approx 6$.  
\item 
In no star is the structure of the
surface velocity field observationally constrained.  For the
understanding of the nature of $\Delta$ frequencies it is of high
interest to compare it to those of the parent variabilities (which 
may also reside in the circumstellar disk).  
\item 
In some stars, there is a mismatch at the $\geq$1\% level between
inferred and calculated frequency differences.
Its significance is not clear, especially since some
of the parent frequencies are members of broad frequency groups.  
\item 
Most $\Delta$ frequencies correspond to periods
ranging from days to weeks.  With 0.2 d, HD 49330 is again an outlier.
Perhaps, this frequency triplet is better described in terms of a sum
frequency.  
\item 
Amplitudes of $\Delta$ frequencies can be both
smaller or larger than those of their parent variabilities.  Only in
$\eta$ Cen is the amplitude of one $\Delta$ frequency more than
twice as large as the amplitude sum of the parent frequencies.  
\item
Stars can exhibit more than one $\Delta$ frequency.  
\item 
One parent frequency can be involved in at least two $\Delta$ 
frequencies. 
\item 
In some stars ($\mu$ Cen [if it does possess
genuine $\Delta$ frequencies], $\eta$ Cen and, perhaps, 28 Cyg)
there is evidence of the active involvement of difference frequencies
in mass loss.  
\item 
With the exception of 25 Ori, mplitude variations during outbursts are not
well documented for $\Delta$ frequencies. In wavelet transforms 
(which are technically biased to such results),
$\Delta$ frequencies may exhibit substantially less amplitude and
phase wobble than other variations (HD 186567).  
\end{list}

\subsection{What is the physical significance of $\Delta$ frequencies?}
With mismatches in frequency of up to a few percent, such
coincidences may in stars with several base frequencies arise by
chance in $\sim10$\% of all cases.  However, if the amplitude is larger
than those of the parent frequencies ($\eta$ Cen, $\mu$ Cen) or there
are links to a variable star-to-disk mass-transfer rate ($\mu$ Cen,
$\eta$ Cen, 28 Cyg), the false-alarm probability becomes negligible.
That is, frequency differences as characterized above are real as a
general phenomenon but individual examples may be spurious.

In pulsating stars, difference ($\Delta$) frequencies are a subset of 
combination
frequencies, which comprise differences, sums, and mixed combinations
of eigenfrequencies.  They are known from white dwarfs
\citep[e.g.,][]{1994ApJ...430..839W}, RR Lyrae
stars \citep[e.g., ][]{2003A&A...398..213M}, 
and $\delta$ Scuti stars \citep[e.g., ][]{1998A&A...331..271B}. In the
rapidly rotating hybrid $\beta$ Cep / SPB star $\beta$ Cen Aa, the
$\Delta$ frequency 0.86 c/d has the highest amplitude of all detected
variations \citep{2016A&A...588A..55P}; for early-type stars in general see
\citet{2015MNRAS.450.3015K}. The theoretical foundation for such nonlinear 
mode coupling was laid by \citet{1982AcA....32..147D}.  However, 
in Be stars circumstellar processes may be involved, which would render the 
two phenomena incomparable.  

As discussed by \citet{2016arXiv160802872R}, rapid rotation adds an
important extra twist: Numerical relations between frequencies in the
observer's frame do not necessarily hold in the corotating frame,
especially considering the change of the latter implied by
differential rotation.  That is, either such numerical relations need
to be looked at with much skepticism or they carry valuable diagnostic
information.  For instance, differential rotation may be one way to
understand the observed imperfect frequency matches.  Perhaps, the
nature of $\Delta$ frequencies may also shed some light on the
unexplained strong spectroscopic preponderance of retrograde modes in
Be stars.

In the context of possible pulsation-assisted mass loss, $\Delta$ 
frequencies may be more relevant than sum frequencies because the much
increased time scales may enable stronger deviations from adiabaticity.
Unfortunately, the repetition time scales of outbursts of most Be
stars are much longer \citep[][$\mu$ Cen is a rare
exception]{2002A&A...393..887M} 
than the lengths
of photometric data strings that can be safely assumed to be free of
long-term instrumental or other artifacts.  More importantly, for a
general explanation of outbursts by $\Delta$ frequencies, they would
need to be present in all Be stars concerned.  As an observational
goal, this is still very far away.



\end{document}